\documentclass[a4paper]{article}
\usepackage{epsfig}
\usepackage{amsmath}
\usepackage{amsfonts}
\usepackage{amssymb}
\usepackage{graphicx}
\usepackage{bm} 
\usepackage{apalike}
\begin{document}

\title{The role of synaptic facilitation in coincidence spike detection}
\author{Jorge F. Mej\'{\i}as and Joaqu\'{\i}n J. Torres\\Departamento de Electromagnetismo y F\'{\i}sica de la 
Materia\\
Universidad de Granada, 18071 Granada Spain}
\maketitle

\begin{abstract}
Using a realistic model of activity dependent dynamical synapses and a standard integrate and fire neuron model we study, both analytically and numerically, the conditions in which a postsynaptic neuron efficiently detects temporal coincidences of spikes arriving at certain frequency from $N$ different afferents. We extend a previous work that only considers synaptic {\em depression} as the most important mechanism in the transmission of information through synapses, to a more general situation including also synaptic {\em facilitation.} Our study shows that: 1) facilitation enhances the detection of correlated signals arriving from a subset of presynaptic excitatory neurons, with different degrees of correlation among this subset, and 2) the presence of facilitation allows for a better detection of firing rate changes. Finally, we also observed that facilitation determines the existence of an optimal input frequency which allows the best performance for a wide (maximum) range of the neuron firing threshold. This optimal frequency can be controlled by means of facilitation parameters.
\end{abstract}

\section{Introduction}

Recently it has been reported that postsynaptic potentials recorded in cortical neurons present dynamical properties depending on the presynaptic activity~\cite{tsodyksPNAS,AVSN97}. This behaviour can be understood by means of the action of different mechanisms taken place at the level of the synapses, as for instance, short-term depression and/or facilitation. The first mechanism considers that the amount of available neurotransmitters in the synaptic button is limited. Thus, the neuron needs some time to recover these synaptic resources in order to transmit the next incoming spike. As a consequence, the dynamics of the synapse is an activity-dependent mechanism producing a non-trivial effect in the postsynaptic response. This picture differs from the classical synaptic description which considers the synaptic strengths as static identities with the only possible time modification due to slow learning processes~\cite{hopfield}. Moreover,
it is well known that short-term depression plays an important role in several
emerging phenomena in the brain such as selective attention~\cite{BT05,MM99}, cortical gain control~\cite{AVSN97}, and complex switching behaviour between activity patterns in neural network models~\cite{torresNC,cortesNC}. However, the complete study of other mechanism present in synaptic transmission on pyramidal neurons, as synaptic facilitation, that compete with depression still is lacking. Synaptic facilitation takes into account that the influx of calcium ions through voltage-sensitive channels favours the neurotransmitter vesicle depletion. This yields to relevant behaviour in synchrony and selective attention~\cite{BT05} and in detection of bursts of action potentials (AP)~\cite{MW00,destexhe04}. Therefore, one would expect that synaptic facilitation have a positive effect in the efficient transmission of temporal correlations between spikes trains arriving from different synapses. 

In this work, we used the phenomenological model of dynamic synapses, including both
depressing and facilitating mechanisms, introduced in~\cite{tsodyksNC} to study the
cooperative effect of both in spike coincidence detection (CD) tasks. That is, we computed the regions, in the space of the relevant parameters, in which a postsynaptic neuron can efficiently detect temporal coincidences of spikes arriving from $N$ different afferents. The aim is to determine the range of the parameters defining the dynamic of the synapses and neuron for which the performance of the neural system under study is improved. Our study shows that facilitation enhances the detection of correlated spikes and firing rate changes in situations for which the mechanism of depression alone does not perform well. These main results are robust and persist even when one decreases the degree of correlation between the afferents. Moreover, synaptic facilitation determines the existence of an optimal frequency which allows the best performance for a wide range of the neuron firing threshold. The location of this optimal frequency can also be controlled by means of facilitation control parameters. This property can be important in neural media constituted by neurons presenting heterogeneity in the firing threshold~\cite{azouz2000} in order to efficiently process information codified, for instance, at this frequency.

\section{The model}
We consider a postsynaptic neuron which receives signals from $N$ presynaptic neurons through excitatory synapses. As a first approximation to model experimental data, we assume that the stimulus received by a particular neuron, as a consequence of the overall neural activity, is modeled by a spike train following a Poisson distribution with mean frequency $f$~\cite{tsodyksNC}. According to the phenomenological model presented in~\cite{tsodyksPNAS}, we consider that the state of the synapse $i$ is governed by the system of equations
\begin{equation}
\begin{array}{lll}
\displaystyle\frac{dx_{i}}{dt}&=&\displaystyle\frac{z_{i}}{\tau_{rec}}-{U}(t)x_{i}\delta(t-t_{sp})
\\
\displaystyle \frac{dy_{i}}{dt}&=&\displaystyle-\frac{y_{i}}{\tau_{in}}+{U}(t)x_{i}\delta(t-t_{sp})
\\
\displaystyle\frac{dz_{i}}{dt}&=&\displaystyle\frac{y_{i}}{\tau_{in}}-\frac{z_{i}}{\tau_{rec}},    
\end{array}
\label{synapse}
\end{equation} 
where $x_{i}$,$y_{i}$,$z_{i}$ are the fraction of neurotransmitters in a recovered, active and inactive
state, respectively. Here, $\tau_{in}$ and 
$\tau_{rec}$ are the inactivation and recovery time constants, respectively. Depressing synapses are obtained for  
${U}(t)=U_{SE}$ constant, which represents the maximum amount of neurotransmitters which can be released (activated) after the arrive of each presynaptic spike. The delta functions appearing in ({\ref{synapse}}) take into account that an AP arrives to the synapse at fixed time $t=t_{sp}$. Typical values of these parameters in cortical depressing synapses are $\tau_{in}=3~ms,$ $\tau_{rec}=800~ms,$ and $U_{SE}=0.5$~\cite{tsodyksPNAS}. 

The synaptic facilitation mechanism can be introduced assuming that ${U}(t)$ has its own dynamics related with the release of calcium from intracellular stores and the influx of calcium from the extracellular medium each time an AP arrives. Here, we consider the dynamics proposed in~\cite{tsodyksPNAS} which assume 
\begin{equation}
{U}(t)\equiv u(t)(1-U_{SE})+U_{SE}
\end{equation}
with
\begin{equation}
\frac{du(t)}{dt}=-\frac{u(t)}{\tau_{fac}}+U_{SE}[1-u(t)]\delta(t-t_{sp}).
\label{udynamics}
\end{equation} 
Here, ${u}(t)$ is a dynamical variable which takes into account the influx of calcium ions into the neuron near the synapse through voltage-sensitive ion channels~\cite{bertramJNEURO}. These ions usually can bind to some acceptor which gates and facilitates the release of neurotransmitters. A typical value for the facilitation time 
constant is $\tau_{fac}=530~ms$~\cite{markramPNAS}. Then, the variable $U(t)$ represents the fraction of neurotransmitters that are being activated, either by the arriving of a presynaptic spike ($U_{SE}$) and by means of facilitating mechanisms ($u(t)(1-U_{SE})$).  

One can think that the postsynaptic current generated in a particular synapse is proportional to the fraction of neurotransmitters which are in the active state, that is, $I_{i}=A_{SE}$ $y_{i}$, where $A_{SE}$ is the maximum postsynaptic current that can be generated~\footnote{Note that the synaptic conductance rather than the synaptic current depends on $A_{SE}\cdot y(t),$ however our assumption for the current is a good approximation when the membrane potential $V(t)$ is below the firing threshold $V_{th}$ and $\tau_{m}\gg \tau_{in},$ so that $V(t)$ remains constant during the time in which the synaptic conductance varies.}. Hereafter, we will choose $A_{SE}\approx 42.5\,pA$ which is within the physiological range and gives an optimal system performance for $V_{th}=13mV,$ which is very near to the mean value threshold measured in some cortical areas~\cite{azouz2000}. Then, the total postsynaptic current generated by signals arriving from the $N$ excitatory synapses
can be written as $I_{total}=\sum_{i=1}^{N} {I_{i}}$. This current generates a postsynaptic membrane potential which we modeled using an integration-and-fire (IF) neuron model, that is
\begin{equation}
\tau_m\frac{dV}{dt}=-V+R_{in}I_{total},
\label{neuron}
\end{equation}
where $R_{in}=0.1 \,G\Omega$ and $\tau_{m}=15~ms$ are, respectively, the input resistance and the membrane time constant. These typical values has been taken also from pyramidal cells~\cite{tsodyksPNAS}. The IF neuron model assumes that, once the membrane potential reaches a certain threshold $V_{th}$ above the resting potential $V_{rest}=0,$ an AP is generated and $V(t)$ is reset to zero. In addition, we assume the existence of a refractory period of $\tau_{ref}=5~ms $  during which $V(t)$ remains to zero after the generation of each postsynaptic AP.

\section{Detection of strongly correlated signals}
\begin{figure}[ht!]
\centerline{\psfig{file=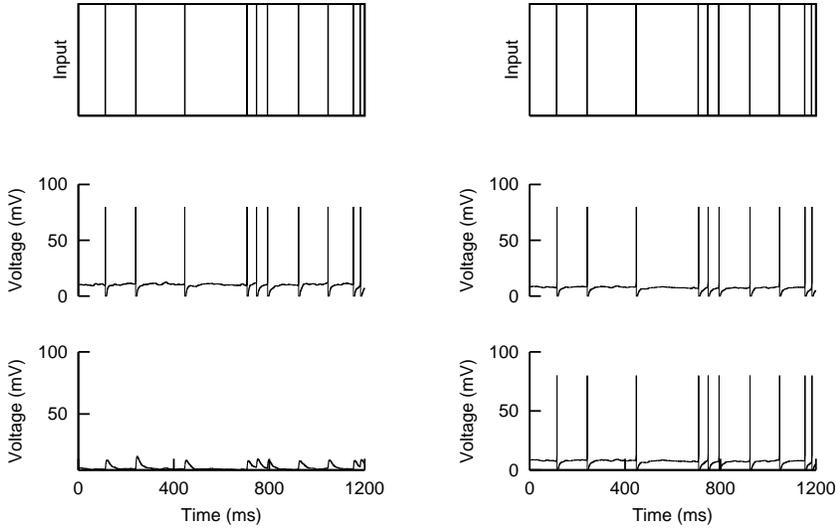,width=12cm}}
\caption{
Response of a postsynaptic neuron receiving Poisson spike train (Top panel) at frequency of $10~Hz$ from $N=1000$ afferents through dynamic synapses. Left and right panels corresponds to the case of depressing and facilitating synapses, respectively.  In these simulation, $U_{SE}$ takes the values $0.5$ (middle panel)
and $0.01$ (bottom panel), respectively, and the threshold is fixed at $13 \,mV$. The figure shows that facilitation enhances CD tasks for relatively low values of $U_{SE}$.}
\label{fig1}
\end{figure}

First, we have studied the postsynaptic response of a neuron receiving input signals from $N=1000$ excitatory synapses, with a subset of $M=200$ synapses stimulated by identical spike trains. These strongly correlated afferents fire spikes simultaneously, so we can consider them as a {\em signal} term. The remaining $N-M$ synapses
receive uncorrelated spike trains which constitute, therefore, a {\em noisy} background of activity superimposed to the signal. We have investigated, both analytically and numerically, spike coincidence detection (CD) experiments. Our interest is to determine the values of the parameters characterizing the neuron and synapse models, for which the postsynaptic neuron can detect the embedding signal (that is, its response is strongly correlated with the input signal). A typical CD experiment is showed in figure \ref{fig1}.
The figure clearly shows the effect of including facilitation compared with the situation in which only depression is considered. For high values of the parameter $U_{SE},$ the system presents a good performance in the CD of the incoming signals, in both cases. However, for  small $U_{SE}$ the detection of signal is improved in the presence of facilitating mechanisms. In fact, when $U_{SE}$ takes low values, the contribution of depression to ${U}(t)$ (which gives the strength of the synapse) becomes irrelevant. Facilitation, however, still contributes to maintain $U(t)$ highly enough to allow a good performance on the CD task. For even smaller values of $U_{SE},$ for instance lower than  $0.007,$ the depressing-facilitating mechanism also fails. In general, given a particular value of $\tau_{fac}$ there exists a critical value of $U_{SE}$ below which the model does not perform well, even including facilitating mechanisms. On the contrary, for higher values of $U_{SE}$ and in the region in which the model has a good performance, the simulations do not show any differences when one includes facilitating mechanisms. The reason is that the facilitation term becomes irrelevant for high values of $U_{SE}$ and depression is the only mechanism contributing to the dynamics. However, still there is a region in the space of relevant parameters --for intermediate values of $U_{SE}$-- where facilitating mechanisms allows for a better performance in the CD tasks.

\begin{figure}[ht!]
\centerline{\psfig{file=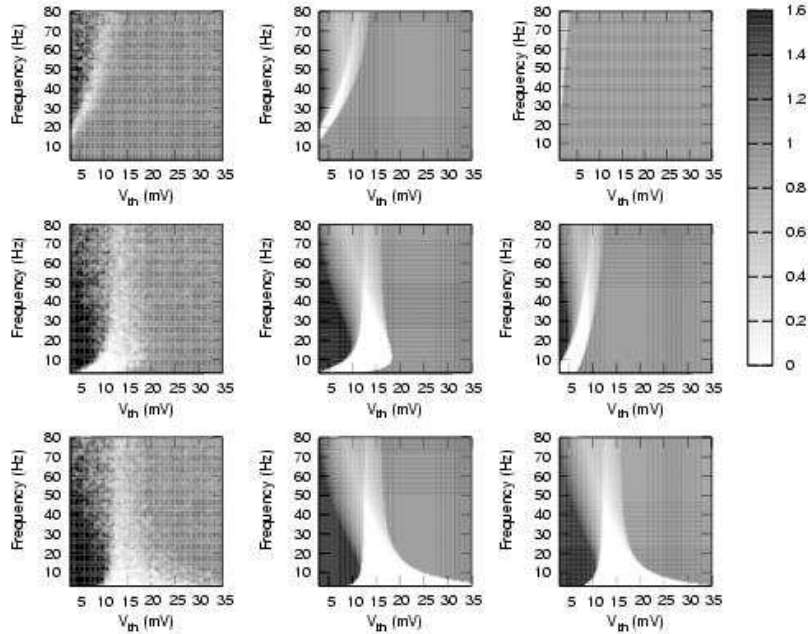,width=11cm}}
\caption{Coincidence detection maps for a system with facilitating synapses. The values of $U_{SE}$ were $0.002$ (top), $0.05$ (middle) and $0.5$ (bottom). One can see that the effect of increasing $U_{SE}$ is the spreading of the region of good CD (light zone) to the right. Simulations (left) confirm the analytical results (center). The right panels represent the same CD regions with only depressing synapses. In all cases $\tau_{fac}$ was $530 \,ms$.}
\label{fig2}
\end{figure}

For a more general evaluation and quantification of the role of the facilitating mechanism, we computed the fraction of errors that occur in the detection of the presynaptic signal by the postsynaptic neuron as a function of the incoming frequency $f$ and the neuron threshold $V_{th}$. These CD error {\em maps} give us a better perspective of the regions where one has a good performance in the space of the relevant parameters. Thus, for each pair $(f,V_{th}),$ we computed in the stationary regime  $1)$ the number of coincidence-input-events in the subset of M coincident afferents, namely $N_{inputs}$, $2)$ the number of output spikes in the postsynaptic neuron occurring immediately within a time-window of $\Delta=5~ms$ after the coincidence-input-events, $N_{hits}$, $3)$ the number of output spikes which are not hits, $N_{falses}$, and $4)$ the number of coincidence-input-events which did not result in output spikes $N_{failures}$ within the time window $\Delta$~\cite{torresDETECTION}.
The fraction of errors is then defined as
\begin{equation}
\%Error\equiv\frac{N_{failures}+N_{falses}}{N_{inputs}}.
\label{error}
\end{equation}
\begin{figure}[ht!]
\centerline{\psfig{file=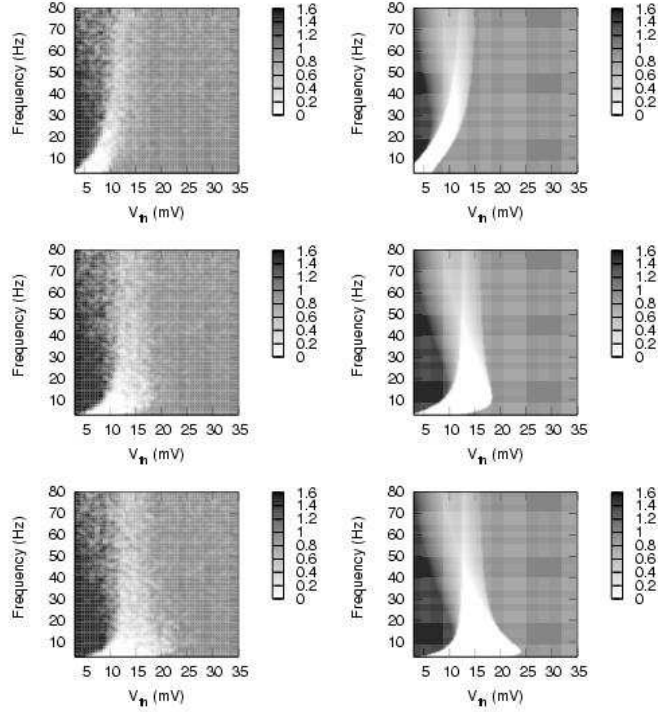,width=9cm}}
\caption{
The error function (as defined in the text) for a system with facilitating synapses. From top to bottom and for fixed $U_{SE}=0.05,$ we consider $\tau_{fac}=50, 530, 1500~ms,$ respectively. The figure shows that the numerically computed error function from simulations (left panels) fits the analytically evaluated (see appendix) error function (right panels). The figure also shows the positive effect of considering stronger facilitation for a better detection of the embedded signal. This is shown by the presence of a larger CD region with a small value for the error function (white areas).}
\label{fig3}
\end{figure}
Analytical expressions for the quantities appearing in (\ref{error}) have been obtained by integration of the model equations (\ref{neuron}-\ref{synapse}) and their derivation is explained in the appendix. 

We have computed both theoretical and numerical CD error functions. The results are showed in figure \ref{fig2}. The light area corresponds to regions where the postsynaptic neuron is able to efficiently detect the coincidence-input-events and to generate a postsynaptic response strongly correlated with the embedded signal. On the other hand, dark areas are regions with a high percentage of errors. These errors can be produced, for instance, when $N_{failures}$ is large, which occurs for $V_{th}$ very large (grey areas), or when $N_{falses}$ increases, normally for small $V_{th}$ in such a way that any current can produce a false event (black areas).
The figure also shows for fixed $\tau_{fac},$ the influence of $U_{SE}$ in the behaviour of the system. Thus, when its value increases (from top to bottom) the width of the light area enlarges and spreads to the right, allowing a better CD for regions with high thresholds. The left panels correspond to numerical simulations whereas the central panels are the same error function evaluated using the analytical formulas, derived in the appendix, into equation (\ref{error}). The figure shows the good agreement between theory and simulations. In the right panels we computed the same regions but considering only the mechanism of synaptic depression. One observes that for only depression and a limited amount of neurotransmitters ($U_{SE}<0.5$), the region for good CD is narrower. In this situation, a large region of good detection is obtained only for $U_{SE}$ near to one. 

Then, we concluded that for the same value of the amount of activated neurotransmitters the overall performance of the system is better with facilitation than if one only consider depressing mechanisms. This conclusion can be also observed when one fixes $U_{SE}$ and varies the facilitation time constant $\tau_{fac},$ as it is shown in figure \ref{fig3}. A large value for  $\tau_{fac}$ means an increase in the duration of the facilitating effect. As a consequence, the region for good detection enlarges compared with the situation of only depression, in special when the fraction of available resources is not too high. 

A detailed observation of figures \ref{fig2} and \ref{fig3} also shows the existence of a certain frequency which allows a good performance for a wide (maximum) range of values of $V_{th}$ (see for instance, middle panels in figure \ref{fig3} that shows a good performance in detecting signal frequencies around $10$ Hz for a threshold ranging from $8$ to $18~mV$). This {\em optimal} frequency decreases as $U_{SE}$ goes to higher values, and becomes zero when only depression mechanism is relevant. Therefore, the presence of facilitation reveals the appearance of an optimal frequency and allows to control it by tuning the facilitation parameters. This result could be important to understand how real neural systems --where different types of neurons may have non-identical firing thresholds-- can self-organize to efficiently detect and process correlated signals.

\begin{figure}[ht!]
\centerline{\psfig{file=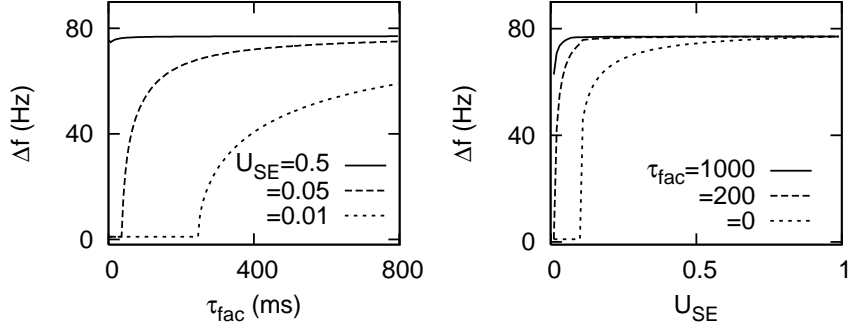,width=12cm}}
\caption{Variation of $\Delta f,$ as defined in the text, for different values of $U_{SE}$ and $\tau_{fac}.$ Left: $\Delta f$ as a function of $\tau_{fac}$ for three fixed values of $U_{SE}.$ Right: $\Delta f$ as a function of $U_{SE}$ for three fixed values of $\tau_{fac}.$ In all cases $V_{th}$ was $13\,m V.$}
\label{fig4}
\end{figure}

\begin{figure}[ht!]
\centerline{\psfig{file=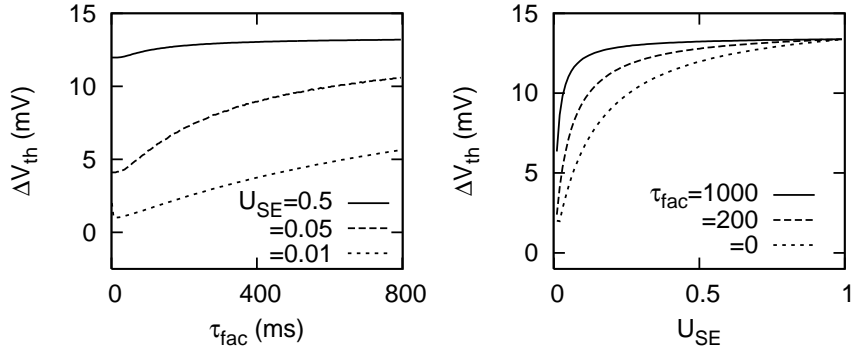,width=12cm}}
\caption{ 
Variation of $\Delta V_{th}$, as defined in the text, for fixed $U_{SE}$ and variable $\tau_{fac}$ (left) and vice versa (right). We can see that the case of $\tau_{fac}=0$ (depression limit) has always a worse behaviour in CD than the facilitation case. The frequency is set to $10\,Hz$.}
\label{fig5}
\end{figure}

Finally, to quantify the variation of the regions of good CD in the presence of facilitation and/or depression, we computed the maximum range of frequencies,  $\Delta f,$  for which the postsynaptic neuron can detect signals with small error (less than $0.5$). We choose a fixed threshold around $V_{th}=13\,m V,$ and study the system behaviour for fixed $U_{SE}$ and varying $\tau_{fac}$ and vice versa. The results are presented in figure \ref{fig4}. The figure shows (left panel) that the range of good CD decreases with $U_{SE}$ for the case of only depressing synapses, and even vanishes for $U_{SE}<0.05.$ However, if facilitating mechanisms are also present the system is able to recover the good performance by increasing the facilitation time constant. The figure also reveals (right panel) that facilitation always enlarges the maximum range of frequencies for any fixed value of $U_{SE}$ (note that the depressing synapses limit is obtained for $\tau_{fac}=0$). Thus, we conclude that for any values of these parameters the inclusion of facilitation in the dynamic of synapses improves the detection towards wider ranges of frequencies. Similar results are found for the maximum range of thresholds which allows good detection of a given frequency $f$, see \ref{fig5}. As a final conclusion, these results show that facilitation expands the regions where the system can detect signals with small errors, both for a wide range of  frequencies and neuron thresholds, and for any possible values of the relevant parameters defining the dynamic of synapses, namely $U_{SE}$ and $\tau_{fac}$.

\section{Effect of jitter}
Detection of coincident signals arriving from different presynaptic neurons have been treated in the previous section in an approximate way, that is, the embedding signal was constituted by fully correlated temporal events. However, in a real situation the incoming signals arriving to a neuron from different synapses are not totally correlated in time. Then and following our previous analysis, if the signal term is formed by presynaptic spikes arriving from $M$ afferents, the presynaptic neurons would never fire exactly at the same time $t_0.$ On the contrary, it is more realistic to consider that the $M$ presynaptic neurons will fire at random times $t_i$ distributed around $t_0$ following, for instance, a Gaussian distribution $p(t_i)$ with a certain deviation or {\em jitter} $\sigma.$ In this section, we consider the implications of this assumption to test the validity of the results previously obtained, and to investigate the effect of the jitter in the detection of signals that are not fully correlated. 
\begin{figure}[ht!]
\centerline{\psfig{file=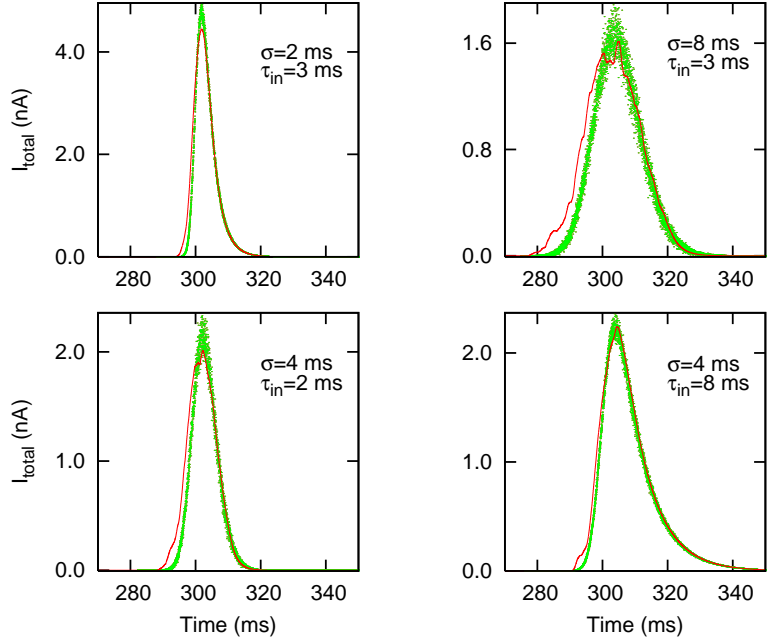,width=12cm}}
\caption{
Excitatory postsynaptic current for a system receiving an single AP from $M=200$ afferents. In each presynaptic neuron the AP occurs at different time $t_i$ which for all neurons is Gaussian distributed around $t_0=300 ms$. The figure shows that the effect of jitter is the spreading of the current curve whereas an increment in the inactivation time constant causes longer right tails. Numerical results (lines) are in concordance with the analytical derivation of the current (dots) (see main text for an explanation).}
\label{fig6}
\end{figure}

We start by computing the excitatory postsynaptic current generated in a synapse $i$ due to a single presynaptic AP occurring at time $t_i,$ that is
\begin{equation}
I_i(t)=I_{peak} \exp[-(t-t_{i})/\tau_{in}] \quad t>t_i
\end{equation} 
where $I_{peak}$ is the steady-state maximum current through a synapse obtained after stimulation with a periodic spike train (see the appendix for details). Since $t_{i}$ is a Gaussian distributed stochastic variable with $\langle t_i\rangle =t_0$ and standard deviation $\sigma$, $q_i(t)\equiv \exp[-(t-t_{i})/\tau_{in}]$ (with $t$ fixed) is also a random variable with range $[0,1]$ and probability distribution given by  
\begin{equation}
{\cal P}[q_i(t)]=\frac{2\tau_{in}}{q_i(t)\;{\rm erfc}\left (-\frac{t-t_0}{\sqrt{2}\sigma}\right )}\frac{1}{\sqrt{2\pi}\sigma}
\exp\left [ -\frac{(t-t_0+\tau_{in}\ln [q_i(t)]^2}{2\sigma^2}\right ]. 
\end{equation} 
where ${\rm erfc}(x)=1-{\rm erf}(x)$ and ${\rm erf}(x)$ is the error function.
One can easily compute the two firsts moments for $P[q_i(t)]$ obtaining:
\begin{equation}
\left\langle q_i(t)\right\rangle_q =\exp\left[\frac{1}{2}(\sigma/\tau_{in})^2-(t-t_0)/\tau_{in}\right] 
\frac{{\rm erfc}\left[ \frac{\sigma^2-(t-t_0)\tau_{in}}{\sqrt{2}\sigma\tau_{in}}\right]}
{{\rm erfc}\left[ -\frac{t-t_0}{\sqrt{2}\sigma}\right]} 
\end{equation} 
\begin{equation}
\left\langle [q_i(t)]^2\right\rangle =\exp\left[2(\sigma/\tau_{in})^2-2(t-t_0)/\tau_{in}\right]
\frac{1+{\rm erf}\left[ \frac{-2\sigma^2+(t-t_0)\tau_{in}}{\sqrt{2}\sigma\tau_{in}}\right]}
{{\rm erfc}\left( -\frac{t-t_0}{\sqrt{2}\sigma}\right)}. 
\end{equation}

In the case of many afferents, the total current in the postsynaptic neuron is $I(t)=I_{peak}\sum_{i=1}^{\nu(t)} q_i(t),$ where $1\le\nu(t)\le M$ is the fraction of the $M$ afferents in which the AP has already generated a postsynaptic response at time $t$, and it is given by $\nu(t)=M\int_{-\infty}^t p(t_i)dt_i.$ This number depends on time due to existence of the jitter that desynchronizes the arriving of the AP in all afferents. Then for $t\ll t_0$  $\nu(t)$ is small, but for $t$ near to and large than $t_0,$  $\nu(t)$ is high and we can use the central limit theorem to obtain:
\begin{equation}
I(t)=I_{peak}\xi(t,t_0), 
\label{eqxi1}
\end{equation} 
where  $\xi(t,t_0)$ is a Gaussian variable with mean and variance given by
\begin{equation}
\left\langle \xi(t,t_0)\right\rangle =\nu(t)\left\langle q(t)\right\rangle 
\label{eqxi2}
\end{equation} 
\begin{equation}
\sigma_{\xi}=\sqrt{\nu(t)[\left\langle q^2(t)\right\rangle -\left\langle q(t)\right\rangle ^2]}
\label{eqxi3}
\end{equation} 
\smallskip
with $\nu(t)=\frac{M}{2}\left[{\rm erf}\left( \frac{t-t_0}{\sigma}\right) +1\right].$ Hereafter we will use this analytical approach to compute CD maps with a jittered signal. 

Since $\nu(t)$ needs to be high in order to use the central limit theorem, one expects that the theoretical current defined by equations (\ref{eqxi1}-\ref{eqxi3}) will fit better the numerical results for $t>t_0.$ This is shown in figure \ref{fig6}, where the analytically computed current after the arriving of M jittered APs (green dots) is compared with the simulated current (red curve), for different values of the jitter $\sigma$ and different values of the inactivation time constant $\tau_{in}$. The figure shows the good agreement between the theoretically and numerically computed currents. Moreover, one observes that, 
the effect of increasing the jitter is the temporal spreading of the current so that the signal influence occurs during a large period of time but with a smaller amplitude. This will cause a small decreasing in the capacity of the system to detect spikes. On the other hand, if we fix the jitter the effect of increasing $\tau_{in}$ is the appearance of longer tails for $t>t_0$, which would be a desirable effect since the response to the next incoming AP will be higher. However, no changes are detected in the amplitude of the current when $\tau_{in}$ is modified. Note that the effect of jitter does not depend on other parameters driving the dynamics of synapses, as $U_{SE},$ $\tau_{rec}$ or $\tau_{fac}$ which only affect to the amplitude $I_{peak}$ (see the appendix). Therefore, one should not expect a strong effect of jitter on the emergent properties due to facilitation and/or depression.

In order to compute CD maps we have to compute the voltage generated by the jittered signal, so that we have to integrate the Langevin equation 
\begin{equation}
\tau_m \frac{dV}{dt}=-V+R_{in}I_{peak}\sum_{t_0}\xi(t,t_0)
\label{langevin}
\end{equation} 

Here the sum extends to a train of {\em events,} each one consisting of $M$ jittered AP centered around a particular instant of time $t_0$ in the event's train. In order to give a first approximation to the solution of this equation, the fluctuations are neglected. Therefore the factor $\xi(t,t_0)$ is now a Gaussian function of time, centered at $t_0$ for each event in the train. Using standard methods and assuming a periodic train of events occurring at $t_0=0,1/f,2/f,\ldots,$ one can easily integrate the equation (\ref{langevin}) obtaining

\begin{equation}
V(t)=\exp(-t/\tau_m)\left[ \exp(-1/f\tau_m)\frac{W(1/f)}{1-\exp(-1/f\tau_m)}+W(t)\right] 
\label{potential}
\end{equation}  

where\begin{equation}
W(t)=\frac{R_{in}}{\tau_m}\int_0^t {\exp(t'/\tau_m)I(t')dt'}
\end{equation} 

\begin{equation}
I(t)=I_{peak}\left\langle \xi(t,0)\right\rangle 
\end{equation} 
which determines the evolution of the membrane potential. Simulation shows that this expression is also valid for Poisson event trains (data not shown). Then, one can use ({\ref{potential}}) to evaluate the CD maps similarly to the case of $\sigma=0$ (non-jittered events). Indeed, as it is shown in the appendix, to do that is necessary the evaluation of the maximum value of $V(t)$ generated by the signal term ($V_m$) during the signal event duration. In the practice, this can be analytically done only in the case of $\sigma=0.$ For $\sigma\neq 0,$ $V_m$ must be numerically computed from (\ref{potential}). 

\begin{figure}[ht!]
\centerline{\psfig{file=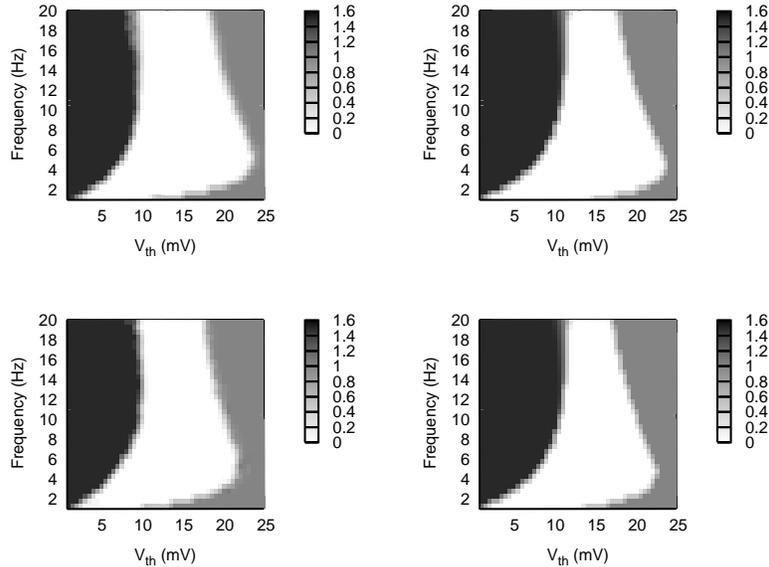,width=13cm}}
\caption{
Coincidence detection maps for the case of arriving presynaptic signals with a certain value of the jitter. A comparison between situations with low ($\sigma=0.5$) and high ($\sigma=3$) jitter is showed (top and bottom panels respectively). One can see that the effect of increasing the jitter is a small and non-relevant decreasing of the good CD region. These results have been found both numerically (left) and analytically (right). The values for the facilitation parameters were $\tau_{fac}=1000~ms$ and $U_{SE}=0.1$.}
\label{fig7}
\end{figure}

The maps for the detection of jittered events are presented in figure \ref{fig7}. An important conclusion is that the CD maps here are qualitatively the same as the maps obtained previously in the zero-jitter case. Increasing the value of the jitter yields to a decreasing of the area of good performance, as one could expect. However, this effect in the light zone is not too dramatic. Indeed, one observes that the shape of the regions does not change too much for two very different values of the jitter, namely $\sigma=0.5$ (top panels) and $\sigma=3$ (bottom panels) (see figure \ref{fig7}). The figure also shows the good agreement between theory (right panels) and simulations (left panels).
In addition, the jitter also causes a small delay on the reaching of the membrane threshold (cf. figure \ref{fig6} where one has the event at $t_0=300~ms$ and the maximum of the generated current occurs at $t=t_0+\delta t$). This fact turns into an increment in the number of failures and false hits. Then, the numerical counting of hits, failures and falses is affected because we consider a fixed detection temporal window of $\Delta=5~ms.$ For $\sigma=3,$ we solved this problem increasing $\Delta$ to $7~ms$ and the result is showed in figure \ref{fig7} (bottom panels). Such effect must, however, be taken into account in situations with a high value of the jitter. 

Finally, the analysis presented in this section concludes that the main results obtained in the previous section are robust for a more realistic treatment of the input presynaptic signals.

\section{Detection of presynaptic firing rate changes}
In the previous study we have considered the overall firing rate as a fixed parameter. This assumption is not realistic and, more interesting, is to consider the presynaptic firing rate as a dynamic variable as it happens in real neuronal tissue. The rate changes during normal functioning of neural systems in the presynaptic current leads to a transient behaviour in the excitatory postsynaptic potential (EPSP) which could cause a burst or an AP in the postsynaptic neuron~\cite{tsodyksPNAS,torresDETECTION,AVSN97}. The question that arises is if the postsynaptic neuron is able to detect synchronous changes (increases) in the afferent firing rates. This property have been found only for depressing synapses and not for static synapses~\cite{torresDETECTION}. Another question is if synaptic facilitation could have some positive effect in the detection of these rates changes by the postsynaptic neuron. In this section we try to answer this question by studying the effect of increasing facilitation in spite of depression in the optimal detection of rate changes in the presynaptic current. 

\begin{figure}[t!]
\centerline{\psfig{file=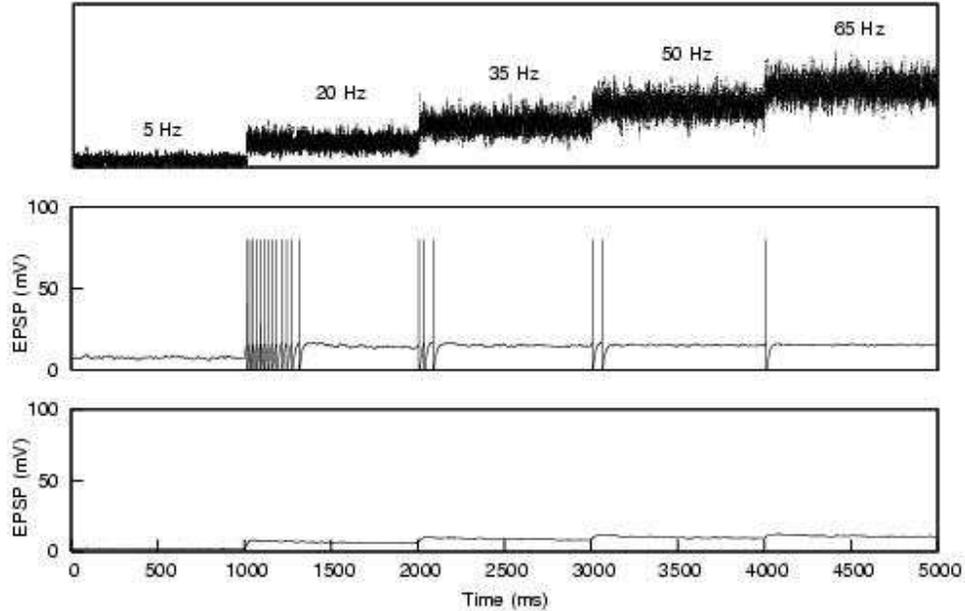,width=13cm}}
\caption{
Detection of firing rate changes with depressing and facilitating synapses. The top panel shows the mean firing rate of the $N=1000$ presynaptic neurons as a function of time. Middle and bottom panels shows the response of the postsynaptic membrane potential for facilitating and depressing cases, respectively. In these simulations parameters were $U_{SE}=0.1$ and $\tau_{fac}=500\,ms(0~ms)$ for the facilitating (depressing) case, respectively. Detection of variations onto lower frequencies are not possible with these synaptic mechanisms.}
\label{fig8}
\end{figure}

To start, we assume a population of $N=1000$ afferents firing uncorrelated Poisson spike trains with a certain frequency $f$ into a postsynaptic neuron. This population changes its mean firing rate every $1000 \,ms$. The figure \ref{fig8} shows a comparison in the output of the postsynaptic neuron for facilitating and depressing synapses. The threshold for firing was fixed in $V_{th}=17\,ms$ and $U_{SE}=0.1$. Simulations show that facilitating synapses ($\tau_{fac}=500\,ms$) allow for a better detection of rate changes, and over a large range of frequencies, than depressing synapses. In general, the regions in which depressing and facilitating synapses perform well can vary. Thus, there are particular situations where facilitation is needed to detect presynaptic rate changes. 

A simple theoretical approach can help us to find the regions of good firing rate changes detections. To obtain such transient behaviour which allows to a rate-change detection, the threshold of the postsynaptic neuron must satisfy $C f_2 \omega(f_1) > V_{th} > C f_2 \omega(f_2)$, where $f_1$ is the initial rate, $f_2$ is the firing rate after the change, $C=R_{in} N \tau_{in}$  and $\omega (f)$ is the stationary postsynaptic current strength for a given frequency. From the system of equations (\ref{synapse}) and following a reasoning similar to the strategy used in the appendix, one can easily obtain
\begin{equation}
\omega(f)=\frac{A_{SE}U_{\infty}}{1+f\tau_{rec}U_{\infty}}
\end{equation} 
where $U_\infty$ is the steady state valued of $U(t)$ (see the appendix).
If now we fix the frequency step $\delta f=f_2-f_1$, the resulting expressions will only depend on $f_1$. Since for large enough frequencies $C f_2 \omega(f_1)$ is a decreasing function of $f_1$ and $C f_2 \omega (f_2)$ is an increasing function of $f_2$ (and therefore of $f_1$), these two tendencies will converge for some $f_1$. This leads to a close area of good rate change detection between the two curves.

\section{Discussion}

In the last years, there was an increasing interest in the study of the computational functionality of synaptic activity dependent processes, as synaptic facilitation and depression, in real systems and neural networks models~\cite{abbott04,destexhe04,torresNC,torresDETECTION}. In particular, recently it has been reported how these processes affect the stability of the attractors of the dynamics and, as consequence, it emerges an oscillatory phase where the activity of the system is continously jumping among the attractors~\cite{torresNC,cortesNC,torresNC2006}. This jumping behavior could explain the emergence of voltage transitions between up and down states observed in cortical areas~\cite{tsodyks06}.

In this paper, we have presented a detailed study of how the competition between synaptic facilitation and depression affects the neural detection of an embedded signal in a background of uncorrelated noise. Our study shows that the inclusion of the facilitation mechanisms enhances the performance of cortical neural systems. In particular, the transmission of information codified in spike trains through the synapses is better and the detection of firing rate changes is also improved compared with the case of only depression. Thus, contrary to which it happens with only depression, the presence of facilitation makes not necessary to have a high value for the maximum amount of active neurotransmitters to efficiently detect correlated signals. This would lead us to think that facilitation has a crucial role in the processing of information through synapses even when the neuron does not have enough synaptic active neurotransmitters. We have also seen that, although important, it is not crucial to have a strong correlation between different presynaptic afferent to have a good detection of signals. Facilitation also determines the existence of an optimal frequency which allows good performance for a wide range of neuron firing thresholds. This result could be important to understand how real neural systems --where different types of neurons with non-identical firing thresholds are connected in a complex way-- can self-organize to efficiently detect and process relevant information~\cite{azouz2000}. Our analysis has also shown that the main conclusions are also valid for realistic jittered signals.

\section{Acknowledgments}
This work was supported by the \textit{MEyC--FEDER} project FIS2005-00791
and the \textit{Junta de Andaluc\'{\i}a} project FQM--165. We thank useful discussion with J. Marro.

\section*{Appendix: Analytical derivation of the error function}

In this section we derived analytical expressions for the functions appearing in the definition of the error function (\ref{error}) used to obtain theoretically the regions for good spike coincidence detection in the $(f,V_{\rm th})$ parameter space. 

First, we assume that the total presynaptic current can be splitted in two terms: a signal term containing the correlated embedded signal and a noise term formed by the background of uncorrelated spikes.

\subsection*{\textbf{Noise contribution}}

To take into account the noise generated by $N-M$ uncorrelated spikes trains, we assume that the current at time $t=t^{*}+\tau$ generated by a single spike arriving to the synapse $i$ at time $t^*$ is given by
\begin{equation}
I_{i}(\tau,t^*)=I_{peak}\exp(-\tau/\tau_{in})
\end{equation}
where $I_{peak}$ represents the averaged stationary EPSC amplitude obtained after stimulation with a periodic spike train, assumption that we also suppose valid for Poisson distributed spike train. After this consideration, one easily obtains from equations (\ref{synapse}-\ref{udynamics}) that
\begin{equation}
I_{peak}=A_{SE}\frac{U_\infty(1-\exp(-1/f\tau_{rec}))}{1-(1-U_\infty)\exp(-1/f\tau_{rec})}
\label{peak}
\end{equation}
with $U_\infty=u_\infty(1-U_{SE}) +U_{SE},$ where $u_\infty$ is the value of $u(t)$ in the stationary state ($t\rightarrow \infty$). For a periodic spike train, $u_\infty$ is given by
\begin{equation}
u_\infty=U_{SE}\frac{\exp(-1/f\tau_{fac})}{1-(1-U_{SE})\exp(-1/f\tau_{fac})}.
\end{equation}
We can compute the mean noise contribution of the current and fluctuations using the standard expressions

\begin{equation}
\begin{array}{ll}
I_{noise}&\equiv\langle I\rangle,\\
\sigma^2_{I_{noise}}&\equiv\langle I^2\rangle-\langle I\rangle^2
\end{array}
\end{equation}
From these definitions and using the central limit theorem we obtain
\begin{equation}
I_{noise}=(N-M)A_{SE}f\tau_{in}U_{\infty}\frac{1-\exp(-1/f\tau_{rec})}{1-(1-U_\infty)\exp(-1/f\tau_{rec})}
\end{equation}
where we assumed that $\tau_{in}\ll \tau_{rec}.$

If we neglect fluctuations ($\sigma_{I_{noise}}=0$), we can write $V_{noise}=R_{in}I_{noise}$. Using this expression one can compute $N_{falses}$ taken into account that false firing occurs when $V_{noise}>V_{th}$ so by a direct integration of equation (\ref{neuron}) in a period of time $T$ gives $N_{falses}\approx T/\{\tau_{ref} -\tau_{m}\ln(1-V_{th}/V_{noise})\}$~\cite{kochbook}. Now using that $f=N_{inputs}/T,$ we finally obtain as in~\cite{torresDETECTION}
\begin{equation}
N_{falses}=\frac{\theta(V_{noise}-V_{th})N_{inputs}}{f(\tau_{ref}-\tau_{m}\ln(1-V_{th}/V_{noise}))}
\label{falses}
\end{equation}
where $\theta(x)$ is the Heaviside step function, which takes into account that for $V_{noise}<V_{th}$ $N_{falses}=0.$ 

To take into account fluctuations of $I_{noise}$ one can use the so called hazard function approximation~\cite{plesser00} but it has been reported that it gives the same results than those obtained using the formula (\ref{falses}) for high frequencies and, on the contrary to the expression (\ref{falses}), it does not work properly for small frequencies~\cite{torresDETECTION}. Therefore, hereafter we will neglect fluctuations in  $I_{noise}$ and use (\ref{falses}) as an approximatively valid expression to analytically compute $N_{falses}.$

\bigskip
\subsection*{{Signal contribution}}
\smallskip

To analyse the signal contribution (arising from M coincident spikes) we used the same method developed in ~\cite{torresDETECTION} for the case of only depressing synapses. That is, assuming that $V(0;t^*)$ is the membrane potential at $t=t^*$ when $M$ coincident spikes arrive, by direct integration of the equation (\ref{neuron}) the membrane potential at time $t=t^*+\tau$ is 
\begin{equation}
V(\tau;t^{*})=e^{\tau/\tau_{m}}\left \{ V(0;t^*) + \frac{R_{in}M I_{peak}}{\tau_{m}\alpha}[e^{\alpha\tau}-1]\right \}
\label{voltage}
\end{equation}
where $\alpha= \frac{\tau_{in}-\tau_{m}}{\tau_{in}\tau_{m}}$ and $I_{peak}$ is given by (\ref{peak}) including all the effects due to synaptic depression and facilitation. If the next signal event (M coincident spikes) occurs at $t=t'$ one can obtain the following recurrence relation:
\begin{equation}
V(0;t')=e^{\Delta t/\tau_{m}}\left \{V(0;t^*) +  \frac{R_{in}M I_{peak}}{\tau_{m}\alpha}[e^{\alpha\Delta t}-1]\right \}
\end{equation} 
with $\Delta t=t'-t^*.$ which allows for computing the stationary value for the membrane potential at the exact time of the signal event arrival (see also \cite{kistler99}), that is:

\begin{equation}
V_{st}=e^{-\Delta t/\tau_{m}}\frac{R_{in}M I_{peak}}{\tau_{m}\alpha}\frac{e^{\alpha \Delta t} -1}{(1-e^{-\Delta t/\tau_{m}})}.
\end{equation} 
We define $V_{signal}$ as the maximum of the membrane potential reached between the arrival of two consecutive signal events separated by a time $\Delta t$. This can be easily computed from equation (\ref{voltage}) with $V(0,t^*)$ replaced $V_{st}$:
\begin{equation}
V_{signal}=\Biggl[\frac{\tau_{m}(1-\exp(-1/f\tau_{m}))}{\tau_{in}(1-\exp(-1/f\tau_{in}))}\Biggr]^{\frac{\tau_{m}}{\tau_{in}-\tau_{m}}}R_{in}MI_{peak}
\end{equation}
where we consider $\tau=\Delta t\backsimeq 1/f.$ 

The expression of $V_{signal}$ allows for evaluate the number of failures assuming that $N_{failures}=N_{inputs}-N_{hits}$. Then, one obtains by direct integration of equation (\ref{neuron}) an using the same reasoning that for $N_{false}$ case that
\begin{equation}
N_{failures}=N_{inputs}\Biggl[1-\frac{\theta(V_{noise}+V_{signal}-V_{th})}{f[\tau_{ref}-\tau_{m}\ln(1-(V_{th}-V_{signal})/V_{noise})]}\Biggr]
\label{failures}
\end{equation}
where we have considered a hit event every time $V_{noise}+V_{signal}$ reach $V_{th}.$ Note that from (\ref{failures}) if $V_{noise}+V_{signal}<V_{th}$ we will have $N_{failures}=N_{inputs}$.

Expression for $N_{falses},$ $N_{failures}$ allows for theoretically compute the number of errors in the CD maps.

\newpage
\bibliography{bibtotalnc}

\end{document}